\pdfoutput=1

\documentclass[11pt]{article}

\usepackage{coling}

\usepackage{times}
\usepackage{latexsym}

\usepackage[T1]{fontenc}

\usepackage[utf8]{inputenc}

\usepackage{microtype}

\usepackage{inconsolata}

\usepackage{graphicx}

\usepackage{algorithm}
\usepackage{algorithmic}
\usepackage{amsmath,scalerel}
\usepackage{subfigure}
\usepackage{booktabs}
\DeclareMathOperator*{\concat}{\scalerel*{\Vert}{\sum}}

\usepackage{amsthm}
\newtheorem{definition}{Definition}
\newtheorem{proposition}{Proposition}

\title{Rumor Detection on Social Media with Temporal Propagation Structure Optimization}

\author{
 \textbf{Xingyu Peng},
 \textbf{Junran Wu\thanks{Corresponding author.}},
 \textbf{Ruomei Liu},
 \textbf{Ke Xu}
\\
 State Key Laboratory of Complex \& Critical Software Environment,
\\
 Beihang University, Beijing, China
\\
 \texttt{\{xypeng, wu\_junran, rmliu, kexu\}@buaa.edu.cn}
}

\begin{document}

\maketitle

\begin{abstract}
Traditional methods for detecting rumors on social media primarily focus on analyzing textual content, often struggling to capture the complexity of online interactions. Recent research has shifted towards leveraging graph neural networks to model the hierarchical conversation structure that emerges during rumor propagation. However, these methods tend to overlook the temporal aspect of rumor propagation and may disregard potential noise within the propagation structure. In this paper, we propose a novel approach that incorporates temporal information by constructing a weighted propagation tree, where the weight of each edge represents the time interval between connected posts. Drawing upon the theory of structural entropy, we transform this tree into a coding tree. This transformation aims to preserve the essential structure of rumor propagation while reducing noise. Finally, we introduce a recursive neural network to learn from the coding tree for rumor veracity prediction. Experimental results on two common datasets demonstrate the superiority of our approach.
\end{abstract}

\section{Introduction}
\begin{figure}[t]
\centering
\includegraphics[width=\linewidth]{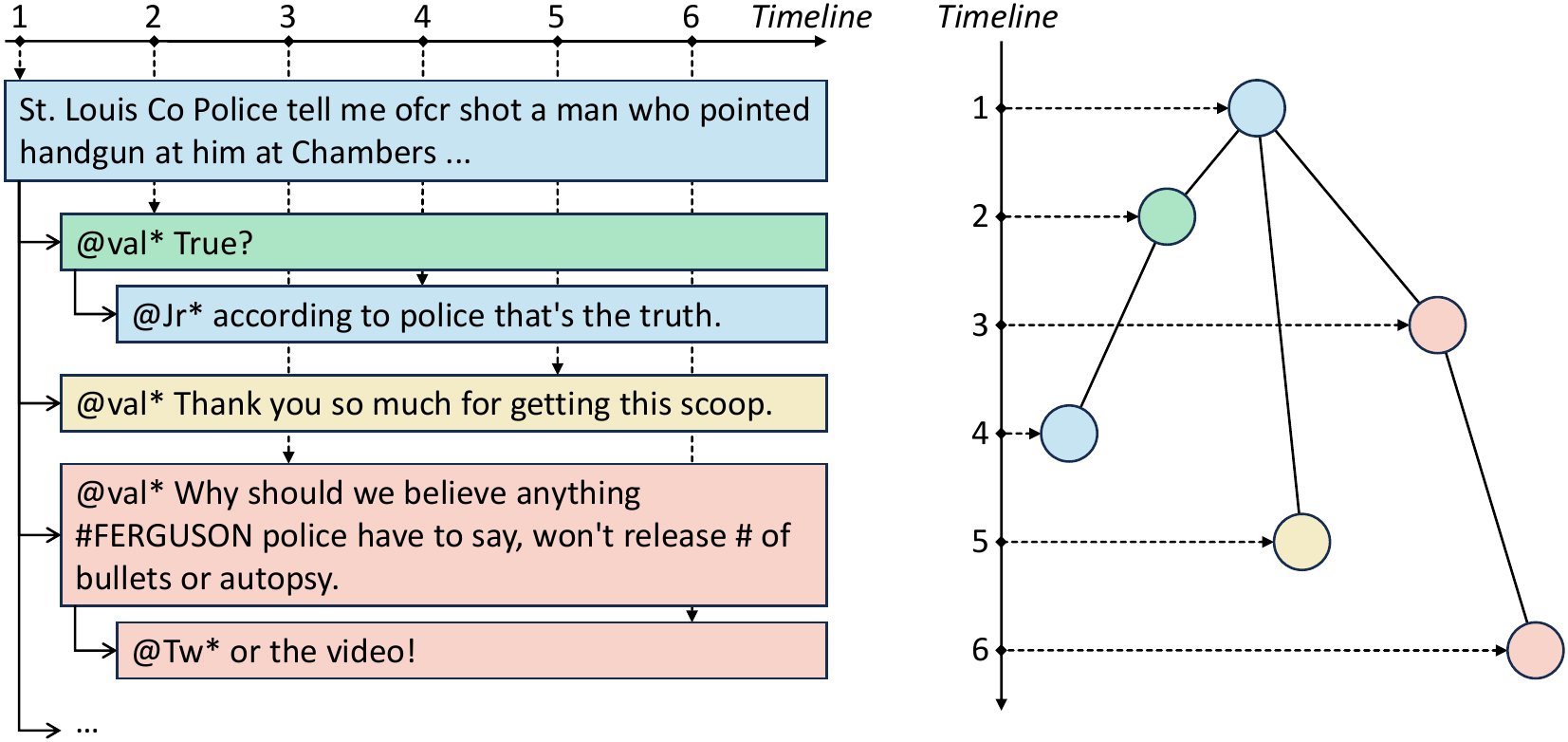}
\caption{Example of a rumor propagation tree related to the \textit{Ferguson} event.}
\label{fig:intro}
\end{figure}

Social media has increasingly become a fertile ground for the generation and dissemination of rumors, which can have significant adverse impacts on society. Detecting rumors on social media is not only a pressing public concern but also a complex and multifaceted challenge. As the flood of false and misleading information continues, extensive efforts have been devoted to automating the process of rumor detection. Traditional methods for identifying rumors often rely on textual analysis \cite{castillo2011information,yang2012automatic,yu2017convolutional}. Recent studies have highlighted the significance of utilizing rumor-spreading patterns, represented as propagation trees, in discerning rumors from non-rumors \cite{ma2018rumor,bian2020rumor}. As illustrated in Figure~\ref{fig:intro}, a propagation tree is formed by a central claim and relevant posts. This tree encapsulates the dynamics of rumor spreading on social media, where the original claim initiates a cascade of reactions and interactions as it disseminates through the network. Therefore, understanding these propagation trees is crucial for effective rumor detection.

Researchers have explored leveraging deep learning models to extract structural representations from propagation trees, employing methods such as Long Short-Term Memory (LSTM) networks \cite{kochkina2017turing}, Recursive Neural Networks (RvNN) \cite{ma2018rumor}, and Transformer networks \cite{khoo2020interpretable}. With the rapid proliferation of Graph Neural Networks (GNNs) in contemporary research, there is a growing trend towards adopting Graph Convolutional Networks (GCN) \cite{bian2020rumor,wei2021towards,wu2024graph} and Graph Attention Networks (GAT) \cite{lin2021rumor,chen2024ssri}. However, these methods tend to overlook the temporal structure inherent in information propagation, despite evidence suggesting its advantages for rumor detection \cite{cheng2021dynamical}. Also, they often assume that the observed structure of the propagation tree is completely accurate, disregarding the presence of potential noise, such as inaccurate relations \cite{wei2021towards}.

\begin{figure}[t]
\centering
\includegraphics[width=\linewidth]{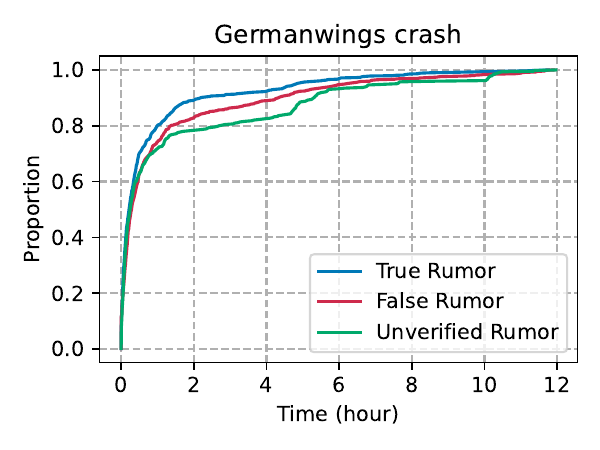}
\caption[]{The Empirical Cumulative Distribution Function (ECDF) plots of the time delay distributions since the initial claim was posted for posts responding to true, false, and unverified rumors in the \textit{Germanwings crash} event\footnotemark{} from the PHEME \cite{zubiaga2016analysing} dataset.}
\label{fig:ecdf_gc}
\end{figure}

The temporal structure of rumor propagation involves the sequence and intervals of posts along the timeline originating from the central claim \cite{huang2020deep}. Figure~\ref{fig:ecdf_gc} shows statistically significant differences in the time delay distributions among the three categories of rumors ($p<0.001$, Anderson-Darling test). \citet{wei2019modeling} observed temporal variations in stance distributions across posts discussing different categories of rumors. These findings underscore the critical need to incorporate temporal information when predicting the veracity of rumors. While previous research \cite{khoo2020interpretable} utilized the time delay of rumor propagation as the transformer's position embedding, it overlooked the hierarchical structure that emerges during rumor propagation. To comprehensively consider both the structural and temporal aspects of rumor propagation, we propose constructing a weighted propagation tree, with each edge assigned a weight corresponding to the time interval between connected posts. This approach facilitates a deeper understanding of the temporal dynamics involved in rumor propagation, including the speed and efficiency of information flow.

\footnotetext{Distributions for more events are available in Figure~\ref{fig:ecdf_all}.}

Drawing upon the theory of structural entropy \cite{li2016structural}, which quantifies the structural complexity and organization of graphs, we derive a coding tree from the propagation tree. The coding tree is a refined version of the original propagation tree that preserves the essential structure of propagation dynamics while reducing noise. In pursuit of the optimal coding tree, we design a greedy algorithm aimed at minimizing structural entropy. To effectively leverage the enriched information within the coding tree, we introduce a recursive neural network for representation learning, referred to as CT-RvNN. CT-RvNN employs an efficient bottom-up message-passing scheme, iteratively propagating information from leaf nodes to the root node. Finally, we create a comprehensive representation of the coding tree through a hierarchical readout strategy, encapsulating critical information for rumor veracity prediction. In comparison to current state-of-the-art methods, CT-RvNN demonstrates superior performance while consuming fewer computational resources. Overall, our contributions can be summarized as follows:
\begin{itemize}
\item We demonstrate the significance of temporal information in predicting the veracity of rumors through statistical analysis. We incorporate temporal characteristics to construct time-weighted propagation trees, capturing the temporal dynamics of rumor propagation.
\item We derive coding trees from the weighted propagation trees through structural entropy minimization. These coding trees preserve the essential structure of rumor propagation while reducing noise. We then employ recursive neural networks to effectively learn rumor representations from these coding trees.
\item Experimental results on two widely used datasets demonstrate the effectiveness and efficiency of our approach. Additionally, in-depth analyses underscore the benefits of incorporating temporal information and the efficacy of the coding tree transformation.
\end{itemize}

\section{Related Work}
\paragraph{Rumor Detection.} Rumor detection on social media has emerged as a prominent and rapidly evolving research field in recent years. Early studies primarily focused on detecting rumors through handcrafted features extracted from various sources, including post contents \cite{castillo2011information}, user profiles \cite{yang2012automatic}, and patterns of information propagation \cite{kwon2013prominent}. \citet{ma2015detect} introduced a time series model designed to capture the temporal evolution of social context information. \citet{wu2015false} and \citet{ma2017detect} employed Support Vector Machines (SVM) with various kernel functions for rumor detection. However, these approaches relied heavily on feature engineering, which is both time-consuming and labor-intensive.

In more recent times, researchers have explored the application of deep learning methods in the domain of rumor detection. \citet{yu2017convolutional} leveraged Convolutional Neural Networks (CNN) to extract essential features from input sequences and capture high-level interactions among these features. \citet{ma2016detecting} adopted Recurrent Neural Networks (RNN) to acquire representations that encapsulate the evolving contextual information of relevant posts over time. \citet{yu2020coupled} divided lengthy threads into shorter subthreads and utilized a hierarchical Transformer framework to learn both local and global interactions among them.

In order to extract informative patterns from both textual content and propagation structures,  \citet{ma2018rumor} introduced an RvNN-based model designed to uncover hidden patterns in tweets organized as propagation trees. \citet{khoo2020interpretable} incorporated a self-attention mechanism to model long-distance interactions within propagation trees. Recent advancements in this field have also seen the incorporation of GNNs. \citet{wei2019modeling} combined structural characteristics and temporal dynamics in the context of rumor propagation using GCN and RNN. \citet{lin2021rumor} represented conversation threads as undirected interaction graphs and employed post- and event-level graph attention mechanisms to extract multi-level rumor-indicative features. \citet{bian2020rumor} focused on both top-down propagation and bottom-up dispersion dynamics among nodes in propagation trees for rumor detection. \citet{wei2021towards} considered the inherent uncertainty in propagation structures, adaptively adjusting weights of unreliable relations to capture robust structural features. \citet{liu2022predicting} proposed utilizing two shared channels for extracting task-invariant textual and structural features, alongside two task-specific graph channels aimed at enhancing structural features. \citet{chen2024ssri} introduced an attention mechanism to capture interaction information among subthreads, incorporating a stance-rumor interaction network that integrates users' stance information with rumor verification. \citet{luo2024joint} employed a graph transformer to concurrently acquire structural and semantic information. Additionally, they utilized a partition filter network to explicitly model rumor- and stance-specific features, as well as shared interactive features. \citet{wu2024graph} developed a multi-view fusion framework that leverages GCN to encode each conversation view, complemented by a CNN to harness consistent and complementary information across all views. Despite the extensive research dedicated to rumor detection, there has been a notable lack of focus on integrating both the structural and temporal aspects of information propagation into the analysis. In addition, the potential influence of noise within the propagation structure has largely been overlooked.

\paragraph{Structural Entropy.} Structural entropy \cite{li2016structural} is an extension of Shannon entropy for structured systems, providing a measure of their structural complexity. In recent years, structural entropy has been successfully applied to various domains, including community structure deception \cite{liu2019rem}, graph classification \cite{wu2022simple}, graph pooling \cite{wu2022structural}, graph contrastive learning \cite{wu2023sega,wu2024uncovering}, graph structure learning \cite{zou2023se}, and text classification \cite{zhang2022hierarchical,zhu2023hitin,zhu2024hill}.

\begin{figure*}[t]
\centering
\includegraphics[width=\linewidth]{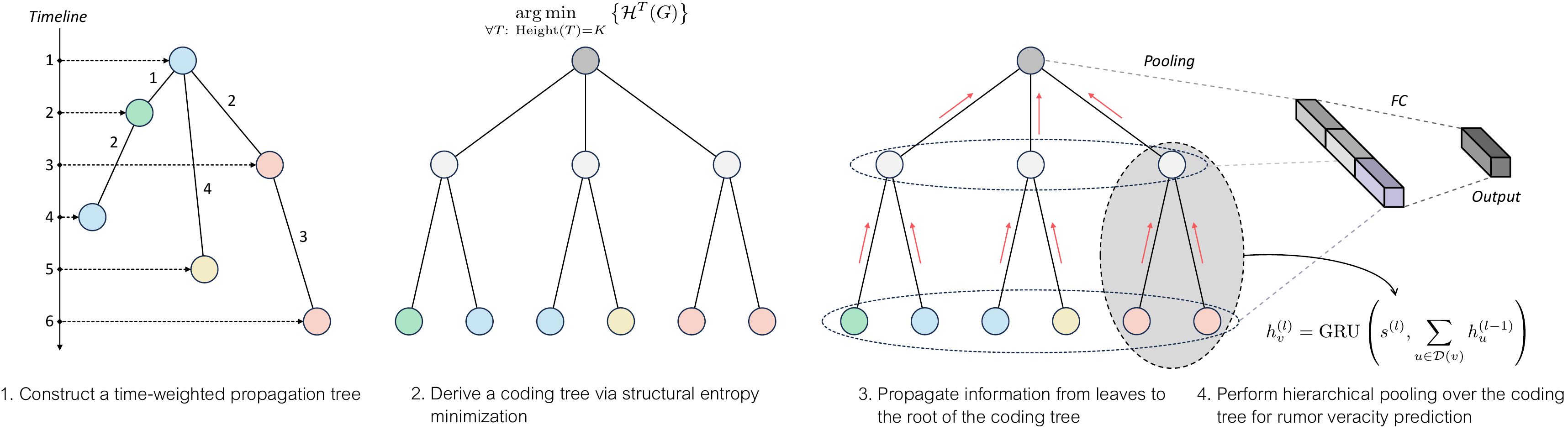}
\caption{Overview of our approach.}
\label{fig:overview}
\end{figure*}

\section{Preliminaries}
\subsection{Problem Statement}
Rumor detection aims to determine the veracity of a claim. Formally, we consider a set of conversation threads on social media platforms, denoted by $\mathcal{C}=\left\{c^1, \ldots, c^{|\mathcal{C}|}\right\}$. Each thread consists of a central claim and a series of relevant posts sorted chronologically: $c^i=\left\{p_1^i, \ldots, p_{n_i}^i\right\}$, where $p_1^i$ is the original claim, and $n_i$ is the number of posts in thread $c^i$. Each post in the thread is represented as $p_j^i=\left(x_j^i, t_j^i, r_j^i\right)$, where $x_j^i$ is the textual content of the post, $t_j^i$ is the publication time of the post, and $r_j^i$ is the index of the post that $p_j^i$ responds to. The propagation process, starting from the original claim, forms a tree-structured graph denoted as $\mathcal{G}^i=\left(\mathcal{V}^i,\mathcal{E}^i\right)$, where $\mathcal{V}^i$ represents the set of posts in thread $c^i$ and $\mathcal{E}^i=\left\{\left(p_{r_j^i}^i, p_j^i\right) \mid j \in [2, n_i]\right\}$ denotes the edges between posts and their responses.

Each claim $c^i$ is associated with a ground-truth label $y^i \in \mathcal{Y}$, which indicates its veracity. The label set $\mathcal{Y}=\{TR, FR, UR\}$ corresponds to the categories of true rumor, false rumor, and unverified rumor, respectively. The primary objective is to learn a classifier $f: \mathcal{C} \rightarrow \mathcal{Y}$ capable of accurately determining the veracity of a claim by analyzing its textual semantics and propagation structure.

\subsection{Structural Entropy}
\citet{li2016structural} introduced the concept of the coding tree of a graph, which serves as a lossless encoding of the graph, with its leaf nodes corresponding to nodes in the graph. They define the structural entropy of a graph $G=\left(V,E\right)$ with respect to a coding tree $T$ as follows:

\begin{equation}
\mathcal{H}^T\left(G\right)=-\sum_{\substack{v_\alpha \in T}} \frac{g_\alpha}{\textit{vol}\left(G\right)} \log _2 \frac{\textit{vol}\left(v_\alpha\right)}{\textit{vol}\left(v_\alpha^{-}\right)}.
\label{eq:se}
\end{equation}

In the equation above, $v_\alpha$ represents a non-root node in the tree $T$, corresponding to a node subset $V_\alpha \subset V$, which comprises the leaf nodes in the subtree of $T$ rooted at $v_\alpha$. $v_\alpha^{-}$ refers to the ancestor of $v_\alpha$ in $T$ and $g_\alpha=\left|E\left(V'_\alpha, V_\alpha\right)\right|$ represents the number of edges from the complement of $V_\alpha$ (denoted as $V'_\alpha$) to $V_\alpha$. In the case of weighted graphs, $g_\alpha$ is the sum of the weights of all the edges between $V_\alpha$ and $V'_\alpha$. The terms $\textit{vol}\left(G\right)$, $\textit{vol}\left(v_\alpha\right)$, and $\textit{vol}\left(v_\alpha^{-}\right)$ refer to the total (weighted) degrees of nodes in $V$, $V_\alpha$, and $V_\alpha^{-}$, respectively.

The structural entropy of the graph $G$ is then defined as the minimum value of $\mathcal{H}^T\left(G\right)$ over all possible coding trees:
\begin{equation}
\mathcal{H}\left(G\right)=\min _T\left\{\mathcal{H}^T\left(G\right)\right\}.
\end{equation}

While identifying the optimal coding tree is important, there are scenarios where decoding a predefined hierarchy for the original graph is more advantageous. In such cases, a coding tree with a predetermined height becomes preferable. To this end, the $K$-dimensional structural entropy is employed to decode the optimal coding tree with a specific height $K$:
\begin{equation}
\mathcal{H}^K\left(G\right) = \min_{\substack{\forall T: \ \textit{Height}\,(T) = K}} \left\{ \mathcal{H}^T\left(G\right) \right\}.
\end{equation}

\section{Methodology}
Figure~\ref{fig:overview} provides an overview of our approach. In Section~\ref{cons}, we leverage the structural and temporal characteristics of rumor propagation to construct time-weighted propagation trees. Next, we devise an algorithm to derive an optimal coding tree from a propagation tree through structural entropy minimization (Section~\ref{cod}). Following this, we introduce our representation learning model, which is designed to acquire meaningful representations of rumors from the coding trees (Section~\ref{rep}). Finally, we classify the veracity of rumors (Section~\ref{tra}).

\subsection{Time-weighted Propagation Tree Construction} \label{cons}
In the context of rumor detection, comprehending the propagation structure of a rumor is critical for assessing its veracity. Following contemporary research practices, our initial step involves constructing a propagation tree. This tree plays a pivotal role in capturing how a claim spreads within a social media platform, with nodes representing the original claim, its subsequent responses, and responses to those responses, collectively forming a hierarchical structure. 

In reality, individuals require time to process and evaluate incoming information before deciding whether to share or endorse a rumor. Therefore, a temporal dimension exists that should be factored into our analysis. To achieve this, we assign weights to the edges of the propagation tree. These weights indicate the time difference between the publication of the connected posts, reflecting the duration it took for the information to propagate from one post to another. Specifically, the weight $w_{jk}$ of an edge connecting a post $p_j$ and its response $p_k$ is calculated as $w_{jk}=t_k-t_j$. As a result, the adjacency matrix of the weighted propagation tree is defined as $\mathcal{A}^i = \left\{w_{jk}^i\right\}^{n_i \times n_i}$, where
\begin{equation}
w_{jk}^i=\left\{\begin{array}{cl}
t_k^i-t_j^i, & \text { if post } p_k^i \text { responses to post } p_j^i, \\
0, & \text { otherwise. }
\end{array}\right.
\nonumber
\end{equation}

\subsection{Coding Tree Construction} \label{cod}
After constructing the time-weighted propagation tree, the next step involves transforming it into a coding tree for structure optimization. To achieve this, we propose a greedy algorithm for minimizing the $K$-dimensional structural entropy, thereby generating the optimal coding tree with a specified height of $K$. To start, we define essential member functions for the tree data structure.

\begin{definition} Given the root node $v_\lambda$ and two of its descendants $v_j$ and $v_k$ in a tree, the function $\textit{join}\,(v_j,v_k)$ inserts a new node $v_\beta$ between the node $v_\lambda$ and the nodes $v_j$ and $v_k$:
\begin{equation}
\begin{aligned}
& v_{\beta}.descendants = \{v_j,v_k\}, \\
& v_{\lambda}.descendants \mathrel{+}= v_{\beta}, \\
& v_{\lambda}.descendants \mathrel{-}= \{v_j,v_k\}.
\end{aligned}
\nonumber
\end{equation}
\end{definition}

\begin{definition} Given an internal node (i.e., a node that is neither the root nor a leaf) $v_\beta$ in a tree, the function $\textit{trim}\,(v_\beta)$ removes the node $v_\beta$ from the tree and integrates its descendants into its ancestor's descendants:
\begin{equation}
\begin{aligned}
& v_\beta.ancestor.descendants \mathrel{+}= v_\beta.descendants, \\
& v_\beta.ancestor.descendants \mathrel{-}= v_\beta.
\end{aligned}
\nonumber
\end{equation}
\end{definition}

\begin{definition} Given a non-root node $v_\alpha$ in a tree, the function $\textit{pad}\,(v_\alpha)$ inserts a new node $v_\beta$ between the node $v_\alpha$ and its ancestor:
\begin{equation}
\begin{aligned}
& v_\beta.descendants = \{v_\alpha\}, \\
& v_\alpha.ancestor.descendants \mathrel{+}= v_\beta, \\
& v_{\alpha}.ancestor.descendants \mathrel{-}= v_\alpha.
\end{aligned}
\nonumber
\end{equation}
\end{definition}

\begin{algorithm}[tb]
\caption{Coding Tree Construction via Structural Entropy Minimization}
\label{alg:algorithm}
\begin{flushleft}
\textbf{Input}: Graph $G=\left(V,E\right)$; coding tree height $K$ \\
\textbf{Output}: Coding tree $T=\left(V_T,E_T\right)$ of height $K$ that minimizes the structural entropy of the graph $G$
\end{flushleft}
\begin{algorithmic}[1] 
\STATE Create a tree $T$ of height 1 with a root node $v_\lambda$ and leaf nodes, where each leaf node corresponds to a node in $V$;

\COMMENT{Expand the tree into a full-height binary tree}
\WHILE{$\left|v_\lambda.descendants\right|>2$}
\STATE Select $v_j$ and $v_k$ from $v_\lambda.descendants$ by $\mathop{\arg\max}_{\left(v_j,v_k\right)}\left\{\mathcal{H}^T\left(G\right)-\mathcal{H}^{T.\textit{join}\,(v_j,v_k)}\left(G\right)\right\}$;
\STATE $T.\textit{join}\,(v_j,v_k)$;
\ENDWHILE

\COMMENT{Compress the tree to the specific height $K$}
\WHILE{$\textit{Height}\,(T)>K$}
\STATE Select an internal node $v_\beta$ from the tree $T$ by $\mathop{\arg\min}_{v_\beta}\left\{\mathcal{H}^{T.\textit{trim}\,(v_\beta)}\left(G\right)-\mathcal{H}^T\left(G\right)\right\}$;
\STATE $T.\textit{trim}\,(v_\beta)$;
\ENDWHILE

\COMMENT{Align the depths of all leaf nodes in the tree to $K$}
\FOR{$v_\alpha \in T$}
\IF{$\textit{Height}\,(v_\alpha^{-})-\textit{Height}\,(v_\alpha)>1$}
\STATE $T.\textit{pad}\,(v_\alpha)$;
\ENDIF
\ENDFOR
\STATE \textbf{return} $T$;
\end{algorithmic}
\end{algorithm}

In Algorithm \ref{alg:algorithm}, we outline the coding tree construction process based on the three functions defined above. Initially, we start with a tree structure that comprises only root and leaf nodes, with each leaf node associated with a node in the propagation tree. We then expand this initial structure into a full-height binary tree using a greedy strategy. This strategy involves iteratively merging two descendant nodes of the root node to form a new division, aiming to minimize structural entropy at each step. Once we have constructed the full-height binary tree, we proceed to compress it to the prescribed height. This compression is achieved using a similar greedy strategy, where internal nodes are iteratively removed while minimizing the increase in structural entropy, until the desired height limit is reached. After the compression process, the resulting coding tree may feature leaf nodes with varying depths. To ensure all leaf nodes attain a consistent depth $K$ for effective representation learning, the function $\textit{pad}$ introduces internal nodes as needed. Proposition \ref{pro} demonstrates that the $\textit{pad}$ operation does not impact structural entropy. Finally, our algorithm yields the optimal coding tree at the specified height, denoted as $T=\left(V_T,E_T\right)$. Here, $V_T^{\left(0\right)}=V$, where $V_T^{\left(0\right)}$ denotes the nodes of height 0 in the coding tree $T$ (i.e., leaf nodes). To facilitate understanding of the coding tree construction algorithm, illustrations of the \textit{join}, \textit{trim}, and \textit{pad} operations, along with the expansion and compression processes, as well as the complexity analysis, are provided in Appendix~\ref{appendix:algo}.

\begin{proposition} \label{pro} For two nodes $v_j$ and $v_k$ in a coding tree $T$ such that $v_j$ is the ancestor of $v_k$, we have $\mathcal{H}^T\left(G\right) = \mathcal{H}^{T.\textit{pad}\,(v_k)}\left(G\right)$.
\end{proposition}
\begin{proof}
    See Appendix~\ref{appendix:proof}.
\end{proof}

\subsection{Rumor Representation Learning} \label{rep}
In this section, based on the temporal information incorporation and the coding tree transformation, we present our model for rumor representation learning. The coding tree serves as a condensed representation of the original propagation structure, preserving essential elements of the rumor-spreading process while reducing data redundancy and noise. Our model is designed to progressively acquire node representations in the coding tree, layer by layer, ultimately culminating in a comprehensive representation of the entire coding tree through a hierarchical readout strategy.

\paragraph{Leaf Node Encoding.} The leaf nodes in the coding tree correspond to posts in the respective conversation thread. We obtain post representations using a sentence encoder:
\begin{equation}
\left\{g_1, \ldots, g_n\right\}=\textit{SentenceEncoder}\left(\left\{x_1, \ldots, x_n\right\}\right),
\end{equation}
where $n$ is the number of posts in the thread. 

Next, we initialize the leaf node representations with the corresponding post representations:
\begin{equation}
h_j^{\left(0\right)}=g_j, \quad \forall j \in \left[1,n\right].
\end{equation}

\paragraph{Tree Positional Encoding.} To capture the hierarchical structure of nodes in the coding tree, we introduce a positional encoding mechanism to enable the model to distinguish nodes at different depths. We define the positional embedding $s^{(l)}$ for nodes at height $l$ as follows:
\begin{equation}
s^{\left(l\right)} = \textit{PositionEncoder}\left(l\right),
\end{equation}
where $\textit{PositionEncoder}\left(\cdot\right)$ generates unique embeddings for each layer of the coding tree.

\paragraph{Bottom-Up Message Passing.}
To facilitate effective learning from the coding tree, we introduce a recursive neural network. This network employs an efficient bottom-up message passing scheme, iteratively propagating information from leaf nodes to the root node. As iterations proceed, the model progressively learns representations for each non-leaf node by aggregating the representations of its descendants, eventually deriving the representation for the root node. In this context, we employ the Gated Recurrent Unit (GRU) \cite{chung2014empirical} as the aggregate function. Consequently, the representation of a non-leaf node at height $l$ in the coding tree is computed as:
\begin{equation}
\begin{aligned}
\bar{h}_v^{\left(l\right)} & = \sum_{u \in \mathcal{D}\left(v\right)} h_u^{\left(l-1\right)}, \\
r^{\left(l\right)} & = \sigma \left(\mathbf{W}_{r}s^{\left(l\right)}+\mathbf{U}_{r}\bar{h}_v^{\left(l\right)}\right), \\
z^{\left(l\right)} & = \sigma \left(\mathbf{W}_{z}s^{\left(l\right)}+\mathbf{U}_{z}\bar{h}_v^{\left(l\right)}\right), \\
\tilde{h}_v^{\left(l\right)} & = \tanh \left(\mathbf{W}_{h}s^{\left(l\right)} + \mathbf{U}_{h}\left(r^{\left(l\right)} \odot \bar{h}_v^{\left(l\right)}\right) \right), \\
h_v^{\left(l\right)} & = \left(1-z^{\left(l\right)}\right) \odot \bar{h}_v^{\left(l\right)} + z^{\left(l\right)} \odot \tilde{h}_v^{\left(l\right)},
\end{aligned}
\end{equation}
where $\mathcal{D}\left(v\right)$ represents the set of descendants of node $v$ in the coding tree, $h_u^{\left(l-1\right)}$ denotes the representation of a descendant node $u$ at the previous layer, and $\odot$ denotes element-wise multiplication. 

\paragraph{Tree Representation Readout.}
The representation $h_T$ of the entire coding tree is generated by combining the representations from each layer in the tree. This is achieved by pooling the representations of nodes at each layer and then concatenating the resulting representations across all layers:

\begin{equation}
h_T=\concat_{l=0}^K\textit{Pool}\left(\left\{h_v^{\left(l\right)} \mid v \in V_T^{\left(l\right)}\right\}\right),
\end{equation}
where $\|$ denotes concatenation, $V_T^{\left(l\right)}$ refers to the nodes at height $l$ in the coding tree $T$, and $\textit{Pool}\left(\cdot\right)$ represents a pooling operation such as summation, averaging, or maximization.

\subsection{Classification and Model Training} \label{tra}
The entire coding tree's representation is fed into a fully connected layer, followed by a softmax function to compute the predicted label probabilities:
\begin{equation}
\hat{y}=\textrm{softmax}\left(\mathbf{W} h_T + \mathbf{b}\right).
\end{equation}

For model training, we adopt the cross-entropy loss as the objective function:
\begin{equation}
\mathcal{L}_\mathcal{C}=-\frac{1}{\left|\mathcal{C}\right|}\sum_{i=1}^{\left|\mathcal{C}\right|} \sum_{j=1}^{\left|\mathcal{Y}\right|} y_j^i \log \hat{y}_j^i+\eta\|\theta\|_2,
\end{equation}
where $y_j^i$ denotes the ground truth probability distribution for the $i$-th sample in $\mathcal{C}$.

\section{Experiments}
\subsection{Experimental Setup}
\paragraph{Datasets.}
We conduct experiments using two publicly available datasets: \textbf{PHEME} \cite{zubiaga2016analysing} and \textbf{Rumoreval} \cite{derczynski2017semeval}. These datasets consist of real-world data collected from Twitter and are widely used in rumor detection. We adopt a leave-one-event-out cross-validation approach for the PHEME dataset, as established in previous studies \cite{wei2019modeling,wei2021towards,liu2022predicting,luo2024joint,chen2024ssri,wu2024graph}. The Rumoreval dataset is evaluated using its official division. Both datasets exhibit imbalances in class distribution. To ensure a comprehensive evaluation of model performance, we prioritize Macro-F1 as the primary evaluation metric and supplement it with Accuracy. Key statistics for both datasets are summarized in Table~\ref{tab:table1}.

\paragraph{Implementation Details.}
To align with prior research \cite{bian2020rumor,wei2021towards,wu2024graph}, we use TF-IDF as the sentence encoder to represent posts as 5000-dimensional vectors based on their textual content. The pooling operation is performed using a summation function. For positional encoding, we employ randomly initialized embeddings, which are further adapted during the model training process. In our experiments on the PHEME and Rumoreval datasets, we set the coding tree height to 7 and 5, respectively.

\paragraph{Baselines.}
In our evaluation, we compare our model against several state-of-the-art rumor detection methods. These methods can be broadly categorized into two groups: \textit{sequence-based methods} and \textit{structure-based methods}. For sequence-based methods, we consider \textbf{BranchLSTM} \cite{kochkina2017turing}, \textbf{HiTPLAN} \cite{khoo2020interpretable}, and \textbf{Hierarchical Transformer} \cite{yu2020coupled}. For structure-based methods, we consider \textbf{TD-RvNN} \cite{ma2018rumor}, \textbf{Hierarchical GCN-RNN} \cite{wei2019modeling}, \textbf{Bi-GCN} \cite{bian2020rumor}, \textbf{ClaHi-GAT} \cite{lin2021rumor}, \textbf{EBGCN} \cite{wei2021towards}, \textbf{STL-GT}\cite{liu2022predicting}, \textbf{MTL-SMI}\cite{liu2022predicting}, \textbf{SSRI-Net}\cite{chen2024ssri}, \citet{luo2024joint}, and \textbf{GMVCN} \cite{wu2024graph}.

\begin{table}[t]
  \centering
    \resizebox{\columnwidth}{!}{\begin{tabular}{lrrrr}
    \toprule
    Method & \multicolumn{2}{c}{PHEME} & \multicolumn{2}{c}{Rumoreval} \\
\cmidrule(lr){2-3} \cmidrule(lr){4-5}          & \multicolumn{1}{c}{Macro-F1} & \multicolumn{1}{c}{Accuracy} & \multicolumn{1}{c}{Macro-F1} & \multicolumn{1}{c}{Accuracy} \\
    \midrule
    BranchLSTM & \multicolumn{1}{c}{0.259} & \multicolumn{1}{c}{0.314} & \multicolumn{1}{c}{0.491} & \multicolumn{1}{c}{0.500} \\
    TD-RvNN & \multicolumn{1}{c}{0.264} & \multicolumn{1}{c}{0.341} & \multicolumn{1}{c}{0.509} & \multicolumn{1}{c}{0.536} \\
    Hierarchical GCN-RNN & \multicolumn{1}{c}{0.317} & \multicolumn{1}{c}{0.356} & \multicolumn{1}{c}{0.540} & \multicolumn{1}{c}{0.536} \\
    HiTPLAN  & \multicolumn{1}{c}{0.361} & \multicolumn{1}{c}{0.438} & \multicolumn{1}{c}{0.581} & \multicolumn{1}{c}{0.571} \\
    Hierarchical Transformer & \multicolumn{1}{c}{0.372} & \multicolumn{1}{c}{0.441} & \multicolumn{1}{c}{0.592} & \multicolumn{1}{c}{0.607} \\
    Bi-GCN\textsuperscript{*} & \multicolumn{1}{c}{0.316} & \multicolumn{1}{c}{0.442} & \multicolumn{1}{c}{0.607} & \multicolumn{1}{c}{0.617} \\
    ClaHi-GAT\textsuperscript{*} & \multicolumn{1}{c}{0.369} & \multicolumn{1}{c}{0.556} & \multicolumn{1}{c}{0.539} & \multicolumn{1}{c}{0.536} \\
    EBGCN\textsuperscript{*} & \multicolumn{1}{c}{0.375} & \multicolumn{1}{c}{0.521} & \multicolumn{1}{c}{0.639} & \multicolumn{1}{c}{0.643} \\
    STL-GT & \multicolumn{1}{c}{0.359} & \multicolumn{1}{c}{0.430} & \multicolumn{1}{c}{0.618} & \multicolumn{1}{c}{0.607} \\
    MTL-SMI\textsuperscript{\textdagger} & \multicolumn{1}{c}{0.409} & \multicolumn{1}{c}{0.468} & \multicolumn{1}{c}{0.685} & \multicolumn{1}{c}{0.679} \\
    SSRI-Net\textsuperscript{\textdagger} & \multicolumn{1}{c}{\underline{0.483}} & \multicolumn{1}{c}{0.568} & \multicolumn{1}{c}{0.743} & \multicolumn{1}{c}{0.750} \\
    \citet{luo2024joint}\textsuperscript{\textdagger} & \multicolumn{1}{c}{0.448} & \multicolumn{1}{c}{0.479} & \multicolumn{1}{c}{\underline{0.754}} & \multicolumn{1}{c}{\underline{0.767}} \\
    GMVCN & \multicolumn{1}{c}{0.441} & \multicolumn{1}{c}{\underline{0.647}} & \multicolumn{1}{c}{0.721} & \multicolumn{1}{c}{0.721} \\
    \midrule
    \textbf{CT-RvNN (Ours)}  & \multicolumn{1}{c}{\textbf{0.486}} & \multicolumn{1}{c}{\textbf{0.794}} & \multicolumn{1}{c}{\textbf{0.792}} & \multicolumn{1}{c}{\textbf{0.786}}  \\
    \bottomrule
    \end{tabular}}%
    \caption{Performance comparison of various methods on the PHEME and Rumoreval datasets. The best result for each metric is highlighted in bold, with the second-best result underlined. Results marked with \textsuperscript{*} are reported in \citet{wu2024graph}. Methods marked with \textsuperscript{\textdagger} are multi-task frameworks that additionally utilize stance information.}
  \label{tab:table2}%
\end{table}%

\subsection{Main Results}
Table \ref{tab:table2} provides an extensive performance comparison of various methods on the PHEME and Rumoreval datasets. Our model consistently outperforms all baseline methods in terms of both Macro-F1 and Accuracy metrics. Specifically, our model achieves a Macro-F1 of 48.6\% and an Accuracy of 79.4\% on the PHEME dataset. For the Rumoreval dataset, it attains a Macro-F1 of 79.2\% and an Accuracy of 78.6\%. A deeper analysis reveals that the notable improvement in Accuracy on the PHEME dataset is primarily driven by enhanced performance in identifying false rumors, which plays a key role in the overall boost.

In comparing various methods, those leveraging GNNs to model conversation structures generally demonstrate superior performance compared to methods relying on sequential models. This superiority stems from GNNs' ability to capture complex dependencies and interactions among nodes in the propagation tree. Notably, Bi-GCN, EBGCN, and GMVCN leverage multiple directed views of conversation threads, integrating both top-down and bottom-up information flows to better capture the underlying patterns of rumor propagation and dispersion. Despite the prowess of these GNN-based methods, our model surpasses them in performance, demonstrating its superior ability to effectively leverage the specific characteristics of rumor propagation. Even when compared to methods that use rumor stance classification as an auxiliary task to enhance rumor veracity prediction, our model still achieves better performance.

Unlike TD-RvNN, which applies the recursive neural network directly to the original propagation tree, our proposed CT-RvNN employs the recursive neural network on the coding tree derived from the time-weighted propagation tree. This adaptation leads to significant performance improvements, underscoring the advantages of incorporating temporal information and the efficacy of the coding tree transformation. We further analyze the specific performance enhancements provided by these two components in the subsequent analysis.

\begin{table}[t]
  \centering
    \resizebox{\columnwidth}{!}{\begin{tabular}{lcccc}
    \toprule
    Method & \multicolumn{2}{c}{PHEME} & \multicolumn{2}{c}{Rumoreval} \\
\cmidrule(lr){2-3} \cmidrule(lr){4-5}          & Macro-F1 & Accuracy & Macro-F1 & Accuracy \\
    \midrule
    CT-RvNN  & \textbf{0.486} & \textbf{0.794} & \textbf{0.792} & \textbf{0.786} \\
    CT-RvNN (Unweighted) & 0.465 & 0.772 & 0.751 & 0.750 \\
    CT-RvNN (Random) & 0.421 & 0.703 & 0.687 & 0.714 \\
    CT-RvNN (Linear) & 0.436 & 0.727 & 0.734 & 0.750 \\
    \bottomrule
    \end{tabular}}%
    \caption{Ablation study results.}
  \label{tab:table3}%
\end{table}%

\subsection{Ablation Study}
In this section, we conduct an ablation study to analyze the impact of various modifications to our model, with the results presented in Table \ref{tab:table3}.

To assess the impact of incorporating temporal weights into the propagation tree, we construct unweighted propagation trees before the coding tree construction and perform experiments, denoted as \textbf{CT-RvNN (Unweighted)}. This modification results in a performance decrease on both datasets. Specifically, the Macro-F1 score decreases to 46.5\% on PHEME and 75.1\% on Rumoreval, while accuracy drops to 77.2\% on PHEME and 75.0\% on Rumoreval. This underscores the significance of considering temporal dynamics in the modeling process.

Next, we investigate the effects of our coding tree construction algorithm. The \textbf{CT-RvNN (Random)} variant randomly generates a coding tree of height $K$. The results show a noticeable reduction in performance, attributed to the random partitioning of nodes during coding tree construction, which disrupts the inherent structural information. In contrast, the coding tree constructed by minimizing structural entropy retains the essential structure of the original propagation tree, thereby enhancing the effectiveness of the learning process.

In the \textbf{CT-RvNN (Linear)} variant, we employ a linear layer to aggregate information from descendants instead of using the GRU. This adjustment aims to investigate the effects of incorporating the GRU for feature aggregation. The Macro-F1 score for this variant is 47.0\% on PHEME and 74.4\% on Rumoreval, with corresponding accuracies of 70.0\% and 75.0\%. Compared to this variant, the GRU-based recursive neural network not only demonstrates superior performance but also exhibits remarkable efficiency advantages through cross-layer weight sharing, especially when the height of the coding tree is high.

\begin{figure}[t]
\centering
\includegraphics[width=\linewidth]{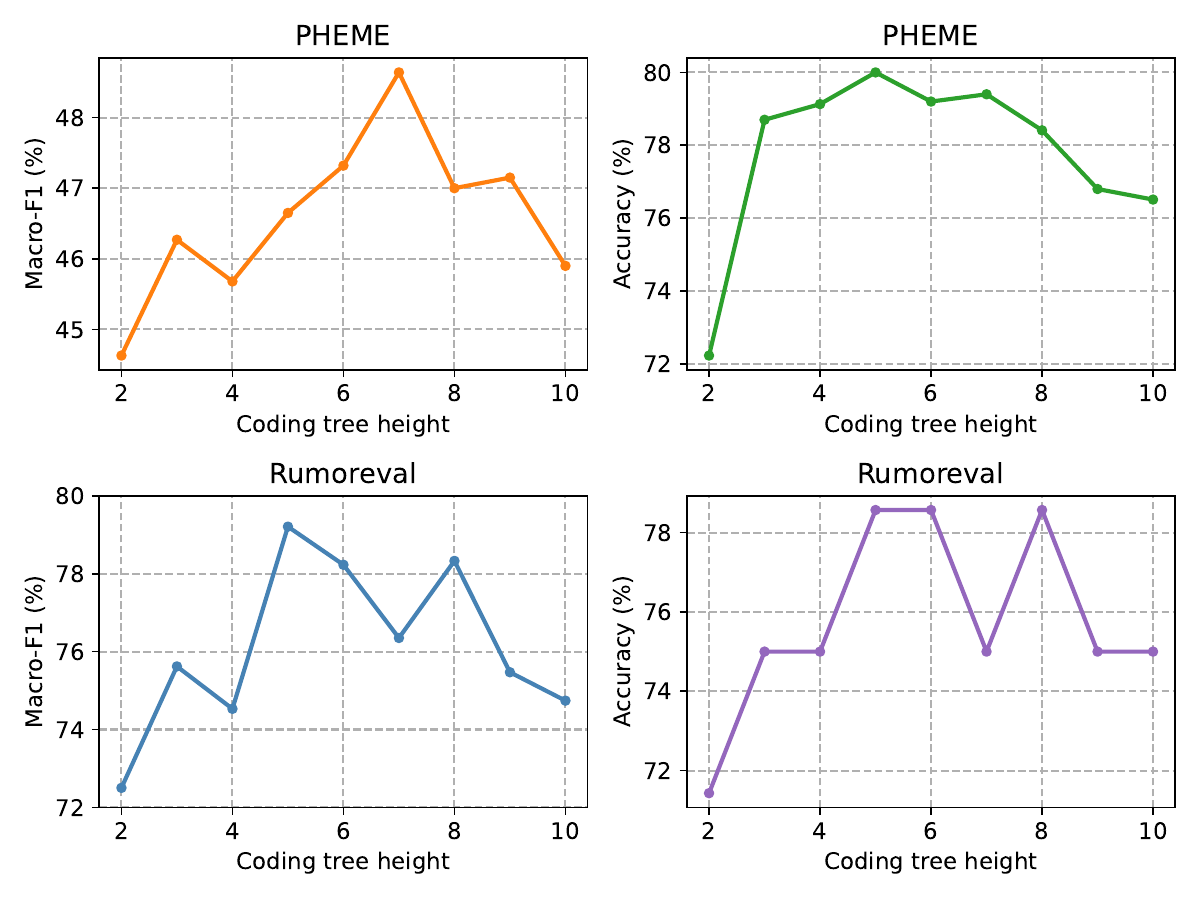}
\caption{Performance analysis results on the PHEME and Rumoreval datasets in terms of coding tree height.}
\label{fig:height}
\end{figure}

\subsection{Performance Analysis}
\paragraph{Impact of Coding Tree Height.}
Coding trees of varying heights capture distinct hierarchical information, thereby influencing the extraction and utilization of information from the leaf nodes. In Figure~\ref{fig:height}, we present an analysis of our model's performance across different coding tree heights on the two datasets. It becomes evident that setting the coding tree height to 2 results in poor performance. This decline in performance can be attributed to the compression of the intricate rumor propagation structure into a coding tree of height 2, which results in the loss of valuable information. To preserve critical details within the propagation structure, a higher coding tree height is essential. For the PHEME dataset, we find that setting the coding tree height to 7 yields the best macro-F1 score of 48.6\%.  However, in terms of accuracy, the optimal coding tree height is 5. Conversely, for the Rumoreval dataset, our model demonstrates optimal performance when the coding tree height is set to 5, achieving a macro-F1 score of 79.2\% and an accuracy of 78.6\%. In summary, selecting an appropriate coding tree height significantly enhances model performance and ensures the effective utilization of information derived from the original claim and the propagation structure.

\begin{figure}[t]
\centering
\includegraphics[width=\linewidth]{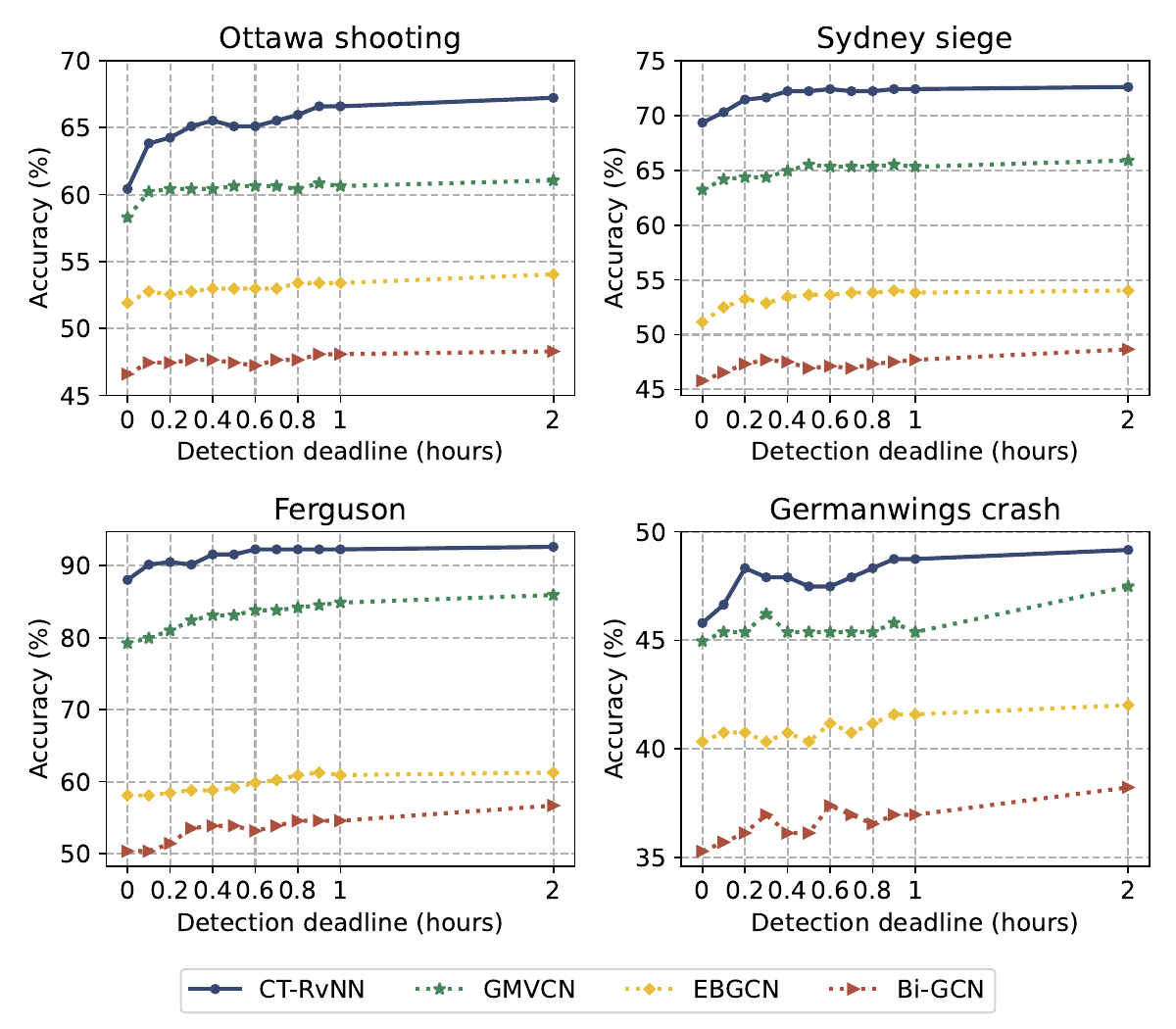}
\caption{Early rumor detection performance analysis results for four distinct events from the PHEME dataset.}
\label{fig:early}
\end{figure}

\paragraph{Early Rumor Detection.}
Early rumor detection aims to identify rumors at their inception, well before they gain widespread traction on social media platforms, enabling timely and appropriate responses. To assess our model's effectiveness in early rumor detection, we follow the approach outlined in \citet{bian2020rumor}. Specifically, we establish various detection deadlines relative to the publication time of the original claim, considering only posts preceding these deadlines to evaluate model performance. Our evaluation is conducted on the PHEME dataset, focusing on folds that use conversation threads related to \textit{Ottawa Shooting}, \textit{Sydney siege}, \textit{Ferguson}, and \textit{Germanwings crash} events for testing. We compare our model against three baseline models: Bi-GCN, EBGCN, and GMVCN. The results of our evaluation are presented in Figure~\ref{fig:early}. 

It is worth noting that the performance of each model generally improves as the detection deadline is extended. This trend is expected, as a longer observation window allows for the accumulation of more available information. Notably, at each deadline, our model, CT-RvNN, consistently outperforms the baseline models. In summary, our model excels not only in long-term rumor detection but also significantly enhances the ability to identify rumors in their early stages.

\begin{figure}[t]
\centering
\includegraphics[width=\linewidth]{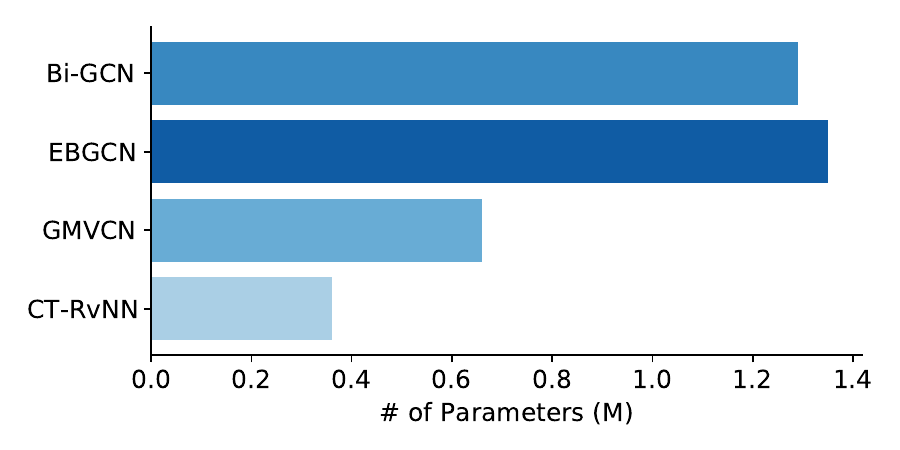}
\caption{Comparison of the parameter counts of various models.}
\label{fig:efficiency}
\end{figure}

\paragraph{Efficiency Analysis.}
In this section, we conduct an efficiency analysis by comparing the parameter counts of various models, including our proposed CT-RvNN and three baseline models: Bi-GCN, EBGCN, and GMVCN. As illustrated in Figure \ref{fig:efficiency}, CT-RvNN exhibits remarkable efficiency in terms of parameter count compared to the baseline models. This showcases its ability to operate with fewer computational resources while still achieving significant performance improvements. Notably, as a recursive neural network founded upon the GRU architecture, CT-RvNN maintains a consistent parameter count regardless of the coding tree height. 

In comparison, Bi-GCN, EBGCN, and GMVCN follow the GCN paradigm. EBGCN and GMVCN, in particular, extend the bidirectional GCN framework introduced by Bi-GCN by constructing two directed graphs to represent rumor propagation: one top-down and the other bottom-up. This dual graph construction results in a doubling of the parameter count. Additionally, Bi-GCN and EBGCN augment the node features in the propagation tree with those of the root node. While this augmentation enhances the utilization of information from the original claim and improves model performance, it also introduces additional parameters.

\section{Conclusion}
In this paper, we use statistical analysis to underscore the significance of temporal information in predicting the veracity of rumors. By leveraging the structural and temporal characteristics of rumor propagation, we construct a time-weighted propagation tree. This tree is then refined into a coding tree through structural entropy minimization, effectively preserving the essential structure of rumor propagation while reducing noise. Finally, we introduce a recursive neural network to learn rumor representation from the coding tree. Experimental results demonstrate that our proposed approach outperforms current state-of-the-art methods while consuming fewer computational resources. Furthermore, in-depth analyses highlight the advantages of incorporating temporal information and the efficacy of the coding tree transformation.

\section*{Limitations}
In line with prior research, we utilize TF-IDF as the sentence encoder. However, the performance of CT-RvNN with alternative sentence encoders, such as static word embeddings, pre-trained language models, or large language models, remains unexplored. 

Recent studies have shown that Transformers can effectively handle tree-structured data. Despite this, we chose to use RvNN over Transformers in this work. While Transformers may offer performance improvements, they introduce significantly more parameters, which conflicts with our primary goal—validating the effectiveness of our method design rather than optimizing performance through a more complex architecture. Additionally, using Transformers would complicate comparisons with prior research that uses RvNN. Nevertheless, integrating Transformers remains a promising direction for future research. 

Moreover, previous studies have highlighted the benefits of incorporating rumor stance classification as an auxiliary task within multi-task learning frameworks to enhance rumor detection. These multi-task frameworks have demonstrated significant improvements over their corresponding single-task ablations \cite{wei2019modeling,yu2020coupled,liu2022predicting,luo2024joint}. Despite consistently outperforming these multi-task models, our CT-RvNN holds untapped potential for further development in multi-task learning. In future work, we plan to integrate stance information into our model to enhance its performance.

\section*{Acknowledgements}
This work has been supported by the Guangxi Science and Technology Major Project, China (No. AA22067070), NSFC (Grant No. 61932002), and CCSE project (CCSE-2024ZX-09). 

\bibliography{custom}

\newpage
\appendix
\setcounter{figure}{0}
\setcounter{table}{0}
\setcounter{equation}{0}
\counterwithin*{figure}{part}
\counterwithin*{table}{part}
\counterwithin*{equation}{part}
\renewcommand{\thefigure}{A\arabic{figure}}
\renewcommand{\thetable}{A\arabic{table}}
\renewcommand{\theequation}{A\arabic{equation}}

\begin{figure}[t]
  \begin{minipage}[t]{0.5\textwidth}
    \centering
    \subfigure[Joining two descendants $v_j$ and $v_k$ of the root node $v_\lambda$.]{
      \includegraphics[width=\linewidth]{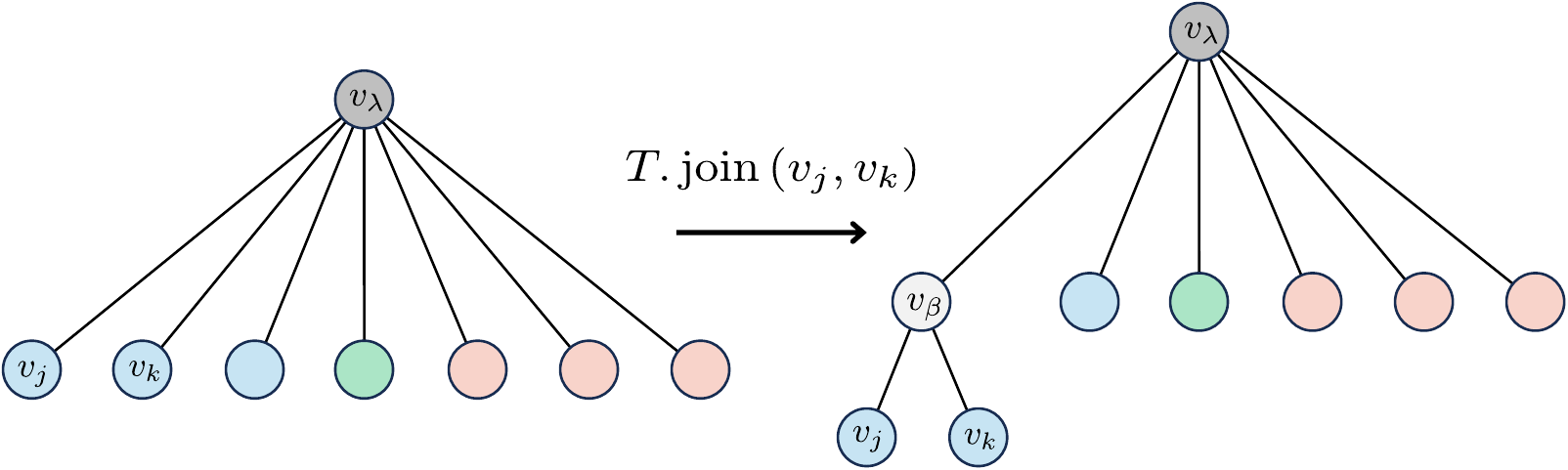}
      \label{fig:joinA}
    }
  \end{minipage}
  \hfill
  \begin{minipage}[t]{0.5\textwidth}
    \centering
    \subfigure[Expanding the initial tree into a full-height binary tree.]{
      \includegraphics[width=\linewidth]{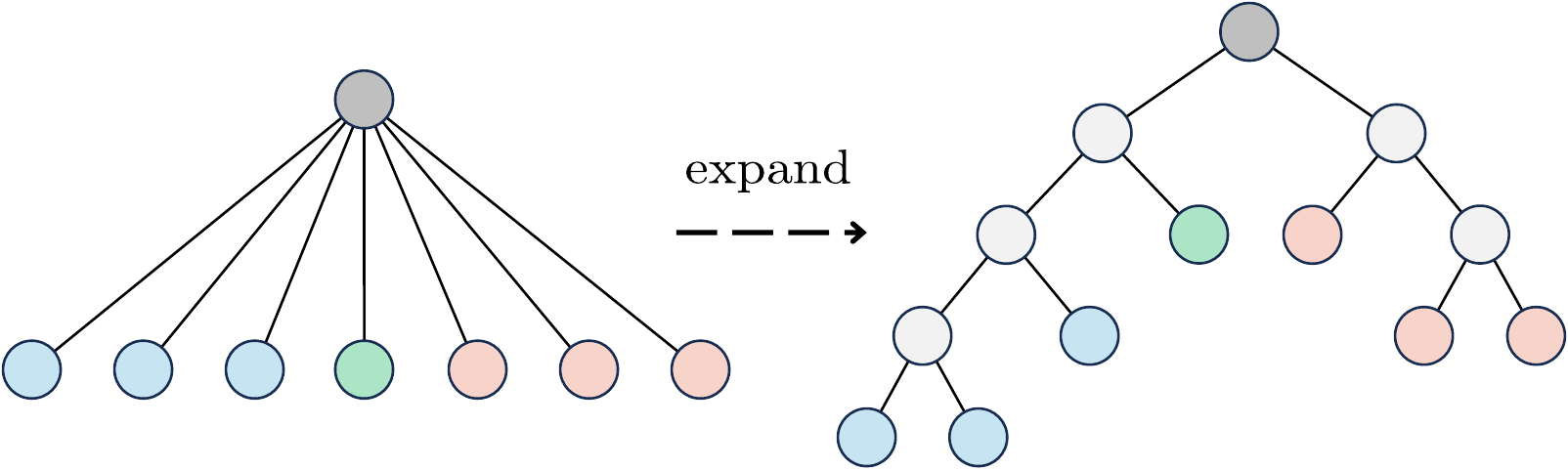}
      \label{fig:joinB}
    }
  \end{minipage}
  \caption{Illustration of the \textit{join} operation and the expansion process.}
  \label{fig:join}
\end{figure}

\section{Supplement to Algorithm~\ref{alg:algorithm}}
\label{appendix:algo}

\textbf{Coding Tree.} A coding tree of a graph $G=\left(V,E\right)$ is defined as a rooted tree $T$ that satisfies the following properties:
\begin{itemize}
    \item Each non-leaf node $v_\gamma \in T$ corresponds to a non-empty subset of $V$, denoted as $V_\gamma$, which comprises the leaf nodes in the subtree of $T$ rooted at $v_\gamma$. Specifically, the root node $v_\lambda$ of $T$ corresponds to the entire set $V$.
    \item Each leaf node $v_\zeta \in T$ corresponds to a unique node in $V$. That is, the subset $V_\zeta$ corresponding to a leaf node $v_\zeta$ is a singleton, containing exactly one node from $V$.
    \item If $v_{\beta_1}, v_{\beta_2}, \cdots, v_{\beta_k}$ are descendants of a non-leaf node $v_\gamma \in T$, then the sets $\{V_{\beta_1}, V_{\beta_2}, \cdots, V_{\beta_k}\}$ form a partition of $V_\gamma$. This means the sets are disjoint and collectively cover all elements of $V_\gamma$.
\end{itemize}

\paragraph{Illustrations.} The \textit{join} operation is illustrated in Figure~A\ref{fig:joinA}. In this operation, a new node $v_\beta$ is inserted between the root node $v_\lambda$ and its two descendants, $v_j$ and $v_k$. As a result, $v_j$ and $v_k$ become the descendants of $v_\beta$, while $v_\beta$ itself becomes a descendant of $v_\lambda$. The selection of $v_j$ and $v_k$ is aimed at minimizing structural entropy. By iteratively selecting nodes $v_j$ and $v_k$ and performing the \textit{join} operation, the initial tree (of height 1) is expanded into a full-height binary tree. This expansion process is illustrated in Figure~A\ref{fig:joinB}, corresponding to Lines 2-5 in Algorithm~\ref{alg:algorithm}.

\begin{figure}[t]
  \begin{minipage}[t]{0.5\textwidth}
    \centering
    \subfigure[Trimming an internal node $v_\beta$.]{
      \includegraphics[width=\linewidth]{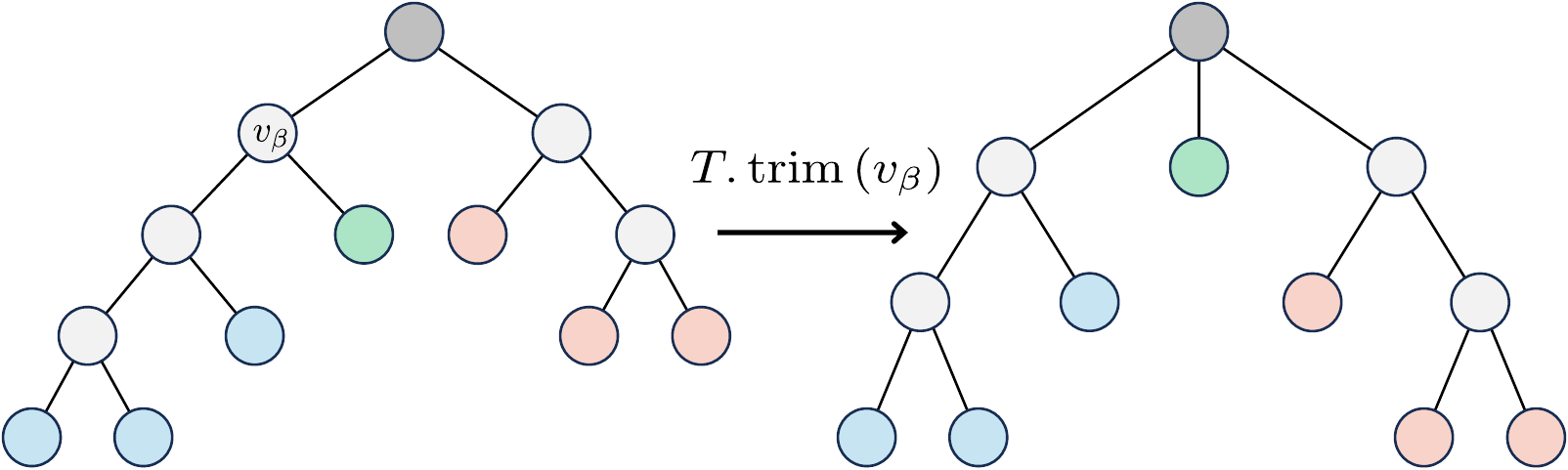}
      \label{fig:trimA}
    }
  \end{minipage}
  \hfill
  \begin{minipage}[t]{0.5\textwidth}
    \centering
    \subfigure[Compressing the tree to a specific height.]{
      \includegraphics[width=\linewidth]{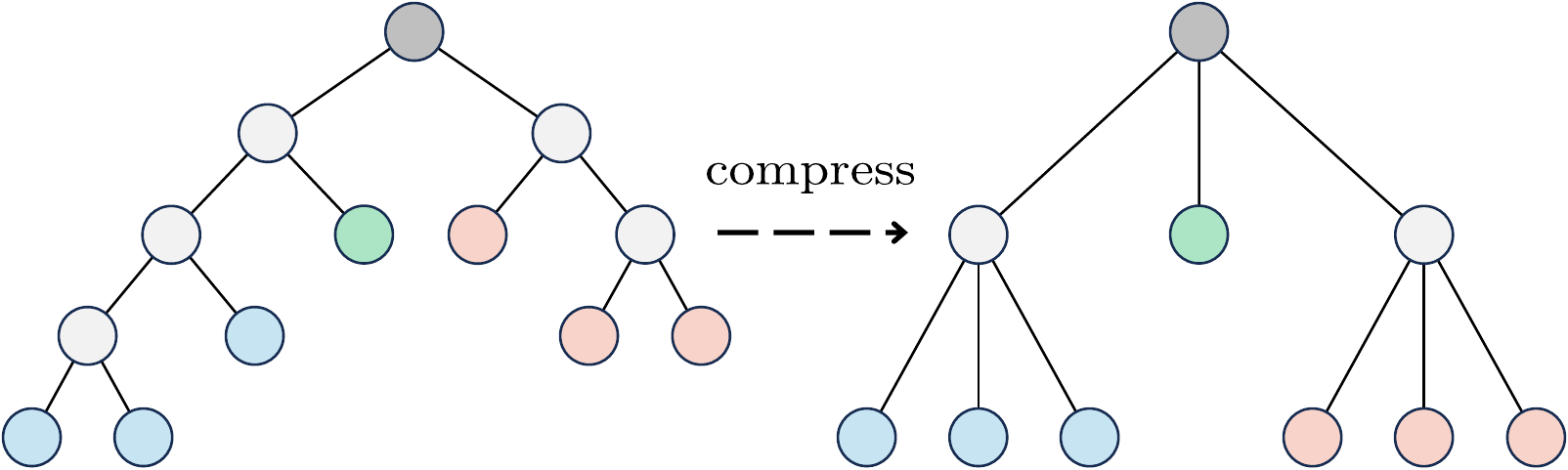}
      \label{fig:trimB}
    }
  \end{minipage}
  \caption{Illustration of the \textit{trim} operation and the compression process.}
  \label{fig:trim}
\end{figure}

The \textit{trim} operation, as illustrated in Figure~A\ref{fig:trimA}, involves removing an internal node $v_\alpha$ from the coding tree and adopting its descendants to its ancestor. The selection of $v_\alpha$ aims to minimize the increase in structural entropy. Through an iterative process of selecting node $v_\alpha$ and executing the \textit{trim} operation, the tree achieves the prescribed height. This compression process is illustrated in Figure~A\ref{fig:trimB}, corresponding to Lines 6-9 in Algorithm~\ref{alg:algorithm}.

\begin{figure}[t]
\centering
\includegraphics[width=\linewidth]{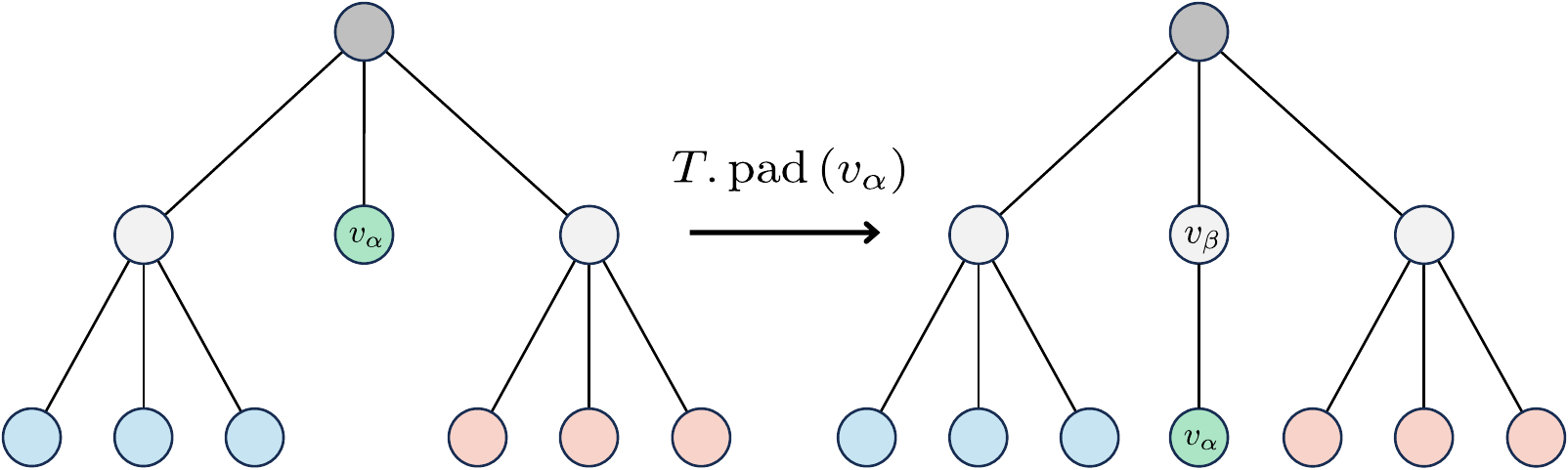}
\caption{Padding a non-root node $v_\alpha$.}
\label{fig:pad}
\end{figure}

The \textit{pad} operation, depicted in Figure~\ref{fig:pad}, entails inserting a new node $v_\beta$ between a non-root node $v_\alpha$ and its ancestor. We iteratively select $v_\alpha$ such that $\textit{Height}\,(v_\alpha^{-})-\textit{Height}\,(v_\alpha)>1$ and apply the \textit{pad} operation until all leaf nodes attain a consistent depth, facilitating the subsequent coding tree representation learning.

\paragraph{Complexity Analysis.} The time complexity of the coding tree construction algorithm is $O\left(2\left|V\right|+h_{\max }\left(\left|E\right| \log \left|V\right|+\left|V\right|\right)\right)$, where $h_{max}$ is the height of the full-height binary tree. It is worth noting that the algorithm tends to construct balanced coding trees, ensuring that $h_{\max}$ is at most $\log \left(\left|V\right|\right)$. Additionally, in typical scenarios, graphs tend to have more edges than nodes, i.e., $\left|E\right| \gg \left|V\right|$, which implies that the runtime of the algorithm scales almost linearly with the number of edges.

\section{Proof for Proposition~\ref{pro}}
\label{appendix:proof}
\begin{proof}
According to Equation~(\ref{eq:se}), $\mathcal{H}^T\left(G\right)$ is the summation of $\mathcal{H}_{v_\alpha}^T=-\frac{g_\alpha}{\textit{vol}\,(G)} \log _2 \frac{\textit{vol}\,(v_\alpha)}{\textit{vol}\,(v_\alpha^{-})}$ for all non-root node $v_\alpha$ in $T$. That is, $\mathcal{H}^{T}\left(G\right)=\mathcal{H}_{v_j}^T+\mathcal{H}_{v_k}^T+\ldots$. Denote the structural entropy after padding $v_k$ as $\mathcal{H}^{T.\textit{pad}\,(v_k)}(G)$. Likewise, $\mathcal{H}^{T.\textit{pad}\,(v_k)}(G)=\mathcal{H}_{v_j^\prime}^{T.\textit{pad}\,(v_k)}+\mathcal{H}_{v_\beta}^{T.\textit{pad}\,(v_k)}+\mathcal{H}_{v_k^\prime}^{T.\textit{pad}\,(v_k)}+\ldots$. Here, $v_\beta$ is the inserted internal node between $v_j$ and $v_k$, while $v_j^\prime$ and $v_k^\prime$ correspond to $v_j$ and $v_k$ after the $\textit{pad}$ operation. The following equations hold:

\begin{align}
\mathcal{H}_{v_k^{\prime}}^{T.\textit{pad}\,(v_k)} & =-\frac{g_k^{\prime}}{\textit{vol}\,(G)} \log \frac{\textit{vol}\,(v_k^{\prime})}{\textit{vol}\,({v_k^{\prime}}^{-})} 
\nonumber
\\
& =-\frac{g_k^{\prime}}{\textit{vol}\,(G)} \log \frac{\textit{vol}\,(v_k^{\prime})}{\textit{vol}\,(v_{\beta})} 
\nonumber
\\
& =-\frac{g_k^{\prime}}{\textit{vol}\,(G)} \log \frac{\textit{vol}\,(v_k^{\prime})}{\textit{vol}\,(v_k^{\prime})} 
\nonumber
\\
& =0, 
\label{eq:1}
\\
\mathcal{H}_{v_{\beta}}^{T.\textit{pad}\,(v_k)} & =-\frac{g_{\beta}}{\textit{vol}\,(G)} \log \frac{\textit{vol}\,(v_{\beta})}{\textit{vol}\,(v_{\beta}^{-})} 
\nonumber
\\
& =-\frac{g_k}{\textit{vol}\,(G)} \log \frac{\textit{vol}\,(v_k)}{\textit{vol}\,(v_j)} 
\nonumber
\\
& =-\frac{g_k}{\textit{vol}\,(G)} \log \frac{\textit{vol}\,(v_k)}{\textit{vol}\,(v_{k}^{-})} 
\nonumber
\\
& =\mathcal{H}_{v_k}^T, 
\label{eq:2}
\\
\mathcal{H}_{v_j^{\prime}}^{T.\textit{pad}\,(v_j)} & =-\frac{g_j^{\prime}}{\textit{vol}\,(G)} \log \frac{\textit{vol}\,(v_j^{\prime})}{\textit{vol}\,({v_j^{\prime}}^{-})} 
\nonumber
\\
& =-\frac{g_j}{\textit{vol}\,(G)} \log \frac{\textit{vol}\,(v_j)}{\textit{vol}\,(v_j^{-})}
\nonumber
\\
& =\mathcal{H}_{v_j}^T.
\label{eq:3}
\end{align}

Thus, we have $\mathcal{H}^T\left(G\right) = \mathcal{H}^{T.\textit{pad}\,(v_k)}\left(G\right)$.
\end{proof}

\begin{figure*}[t]
\centering
\includegraphics[width=\linewidth]{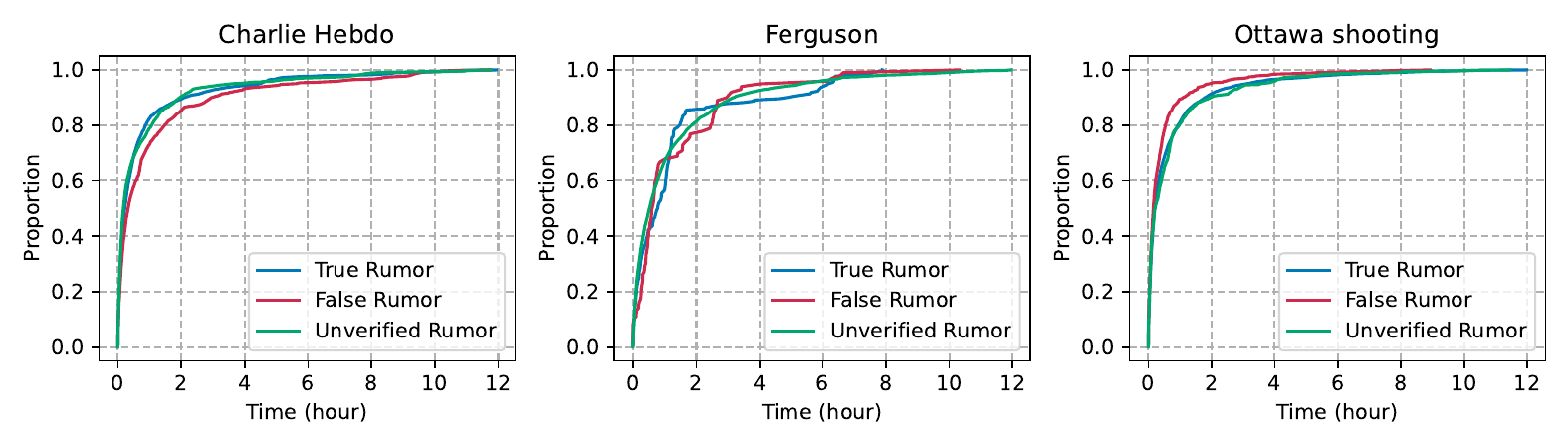}
\caption{The ECDF plots of the time delay distributions since the initial claim was posted for posts responding to true, false, and unverified rumors for \textit{Charlie Hebdo}, \textit{Ferguson}, and \textit{Ottawa shooting} events from the PHEME dataset.}
\label{fig:ecdf_all}
\end{figure*}

\begin{table}[t]
  \centering
    \resizebox{\columnwidth}{!}{\begin{tabular}{lrr}
    \toprule
    Statistic & \multicolumn{1}{c}{PHEME} & \multicolumn{1}{c}{Rumoreval} \\
    \midrule
    \# of users & 18813 & 3859 \\
    \# of posts & 32925 & 5568 \\
    \# of authors & 1024  & 203 \\
    \# of claims & 2402  & 325 \\
    \# of true rumors & 1067  & 145 \\
    \# of false rumors & 638   & 74 \\
    \# of unverified rumors & 697   & 106 \\
    Avg. \# of posts & 13.7  & 17.1 \\
    Avg. depth of the propagation tree & 2.8   & 3.4 \\
    Avg. rumor lifespan (hours) & 8.7   & 23.3 \\
    \bottomrule
    \end{tabular}}%
  \caption{Statistics of datasets.}
  \label{tab:table1}%
\end{table}%

\section{Detailed Experimental Setup}
\paragraph{Datasets.} The PHEME dataset comprises 2402 conversation threads related to nine events. To ensure robust results, we adopt a leave-one-event-out cross-validation approach following established practices. In each fold of the cross-validation, we use the conversation threads associated with one event for testing, while the conversation threads related to the remaining eight events are used for training. 

The Rumoreval dataset contains 325 conversation threads, officially divided into training, validation, and test sets. These conversation threads are related to ten events, and the test set covers two events that are not present in the training and validation sets.

Each claim in both datasets is categorized into one of three classes: true rumor, false rumor, or unverified rumor. 

\paragraph{Implementation Details.} For model training, we use the AdamW optimizer and implement a linear learning rate scheduler with a 6\% warmup and a maximum learning rate of 0.001. Additionally, we apply an L2 regularization weight penalty of 0.0005 to mitigate potential overfitting. 

\paragraph{Baselines.}
In our evaluation, we compare our model against the following competitive baselines:
\begin{itemize}
\item \textbf{BranchLSTM} \cite{kochkina2017turing} is an LSTM-based sequential model designed to capture the conversation structure of a claim and its associated posts.
\item \textbf{TD-RvNN} \cite{ma2018rumor} is a top-down tree-structured model that utilizes recursive neural networks to model the propagation layout of rumors.
\item \textbf{Hierarchical GCN-RNN} \cite{wei2019modeling} utilizes a GCN to encode conversation structures for stance classification and an RNN to capture the temporal dynamics of stance evolution for veracity prediction.
\item \textbf{HiTPLAN} \cite{khoo2020interpretable} is a structure-aware self-attention network that incorporates propagation structural information into the Transformer model.
\item \textbf{Hierarchical Transformer} \cite{yu2020coupled} generates conversation thread representations by dividing each long thread into shorter subthreads and employing BERT to separately represent each subthread.
\item \textbf{Bi-GCN} \cite{bian2020rumor} applies two GCNs on the top-down directed graph and the bottom-up directed graph to learn patterns of rumor propagation and dispersion.
\item \textbf{ClaHi-GAT} \cite{lin2021rumor} employs a claim-guided hierarchical attention mechanism at both post- and event-level to attend to informative posts.
\item \textbf{EBGCN} \cite{wei2021towards} extends Bi-GCN by handling uncertainty in the propagation structure with a Bayesian method, adaptively adjusting weights of unreliable relations.
\item \textbf{STL-GT} \cite{liu2022predicting} utilizes two shared channels to extract task-invariant textual and structural features.
\item \textbf{MTL-SMI} \cite{liu2022predicting} extends STL-GT by incorporating two task-specific graph channels for multi-task learning.
\item \textbf{SSRI-Net} \cite{chen2024ssri} utilizes an attention mechanism to capture interaction details among subthreads and a stance-rumor interaction network to integrate users' stance information with rumor verification.
\item \citet{luo2024joint} employs a graph transformer to concurrently gather structural and semantic information and a partition filter network to explicitly model rumor- and stance-specific features.
\item \textbf{GMVCN} \cite{wu2024graph} is a multi-view fusion framework that utilizes a GCN and a CNN to encode and capture complementary information from various conversation views.
\end{itemize}

\section{Explanations for Figures~\ref{fig:ecdf_gc} and~\ref{fig:ecdf_all}}
In Figures~\ref{fig:ecdf_gc} and~\ref{fig:ecdf_all}, statistically significant differences consistently appear in the time delay distributions among the three categories of rumors. Specifically, noteworthy disparities are evident in the distributions of true and false rumors across all four events. During the \textit{Charlie Hebdo} and \textit{Ottawa shooting} events, the distribution of unverified rumors closely resembles that of true rumors. However, in the remaining two events, the distribution of unverified rumors diverges significantly from the other two distributions. This observation highlights the critical role of temporal information in predicting the veracity of rumors.

\end{document}